\documentclass[aps,prbl,superscriptaddress,reprint,showpacs]{revtex4-2} 
\usepackage{amsmath,amssymb,amsfonts,stmaryrd,wasysym,graphicx,multirow,textcomp,subfigure}

\usepackage[utf8]{inputenc}
\usepackage[version=4]{mhchem}
\usepackage{subfigure}
\usepackage{listings}
\usepackage{siunitx}
\usepackage[english]{babel}
\usepackage{float}
\usepackage[colorlinks=true,citecolor=blue,urlcolor=blue]{hyperref}
\usepackage{bm}
\usepackage{mathtools, nccmath}
\usepackage{lipsum}
\usepackage{amsmath,amsfonts,amssymb,mathtools}
\usepackage{graphicx,float}
\usepackage[ruled,vlined]{algorithm2e}
\usepackage{algorithmic}
\usepackage{minted}
\usepackage{natbib}
\usepackage{booktabs}
\usepackage{siunitx}
\usepackage{dcolumn}
\newcolumntype{d}[1]{D{.}{.}{#1}} 

\begin{document}

\title{Non-uniform grids to accelerate electron-phonon coupling calculations for superconductivity}

\author{Kang Wang}
\affiliation{Department of Materials Science and Metallurgy, University of Cambridge, 27 Charles Babbage Road, Cambridge CB3 0FS, United Kingdom}
\affiliation{Department of Physics, University of Oxford, Parks Road, Oxford, OX1 3PU, United Kingdom}

\author{Siyu Chen}
\affiliation{Department of Materials Science and Metallurgy, University of Cambridge, 27 Charles Babbage Road, Cambridge CB3 0FS, United Kingdom}
\affiliation{TCM Group, Cavendish Laboratory, University of Cambridge, J. J. Thomson Avenue, Cambridge CB3 0HE, United Kingdom}
\affiliation{European Theoretical Spectroscopy Facility, Institute of Condensed Matter and Nanosciences, Université catholique de Louvain, Chemin des Étoiles 8, B-1348 Louvain-la-Neuve, Belgium.}

\author{Sun-Woo Kim}
\affiliation{Department of Materials Science and Metallurgy, University of Cambridge, 27 Charles Babbage Road, Cambridge CB3 0FS, United Kingdom}
\affiliation{Department of Physics, Hanyang University, Seoul 04763, Republic of Korea}

\author{Kieran Bozier}
\affiliation{Department of Materials Science and Metallurgy, University of Cambridge, 27 Charles Babbage Road, Cambridge CB3 0FS, United Kingdom}

\author{Chris Pickard}
\affiliation{Department of Materials Science and Metallurgy, University of Cambridge, 27 Charles Babbage Road, Cambridge CB3 0FS, United Kingdom}
\affiliation{Advanced Institute for Materials Research, Tohoku University, Sendai 980-8577, Japan}

\author{Bartomeu Monserrat}
\email{bm418@cam.ac.uk}
\affiliation{Department of Materials Science and Metallurgy, University of Cambridge, 27 Charles Babbage Road, Cambridge CB3 0FS, United Kingdom}

\begin{abstract}
Efficient and accurate evaluation of electron-phonon interactions is essential for predictive superconductivity calculations, but dense Brillouin zone sampling requirements pose a major computational bottleneck. To overcome this, we introduce an accelerated, highly scalable framework that combines a double delta truncation scheme with non-uniform Farey grids in both $\mathbf{k}$ and $\mathbf{q}$ space. The truncation strategy significantly reduces computational cost by restricting integration to states satisfying energy constraints near the Fermi level, while the Farey grids dynamically resolve sampling errors from localized scattering and stabilize convergence. Applied to elemental aluminium and several high-pressure hydrides, this combined methodology reduces computational cost by up to two orders of magnitude and yields electron-phonon coupling constants and superconducting critical temperatures ($T_{\mathrm{c}}$) typically within about 5\% of full-grid benchmarks. This approach overcomes sampling bottlenecks associated with dense uniform grids, making accurate, low-cost predictions of superconductivity in complex quantum materials feasible.

\end{abstract}

\maketitle

\section{Introduction}

Electrons in solids are coupled to phonons, and this electron-phonon (e-ph) coupling governs a wide range of material properties\,\cite{Grimvall1981,Giustino2017,Claes2025}, such as electrical resistivity\,\cite{Park2014,Bernardi2014,Li2015,Liu2017,Jhalani2017,Ma2018,Zhou2018PRL,Sohier2018,Zhou2019,Lee2020}, carrier mobility\,\cite{Ponce2018,Zhou2016}, charge density waves\,\cite{Xie2022,Luo2022,Wang2024}, and conventional superconductivity\,\cite{BCS1957,McMillan1968,AllenDynes1975}. Among these, conventional superconductivity is particularly striking, as e-ph coupling underpins the emergence of a macroscopic quantum state. Microscopically, phonon-mediated attraction leads to the pairing of electrons into Cooper pairs, which is the cornerstone of the Bardeen-Cooper-Schrieffer (BCS) theory\,\cite{BCS1957}. A quantitative description of e-ph coupling is therefore crucial for both understanding many known superconductors and for guiding the discovery of new ones.

Calculations of electron-phonon coupling for superconductivity can be performed using either analytical models\,\cite{BCS1957,Bardeen1950,Herring1956,Frohlich1954,Mahan2000,Ganose2021} or first-principles approaches\,\cite{Baroni2001,Margine2013,Giustino2017,Monserrat2018,Duan2019,Luo2024PRX,Sanna2020,Waldecker2016,Wang2025PRXEnergy}. Analytical models rely on a small number of parameters to describe the interaction phenomenologically; first-principles methods, typically based on density functional perturbation theory (DFPT)\,\cite{Baroni2001}, offer a fully \textit{ab initio} framework for quantitatively evaluating e-ph coupling. Within DFPT, the key quantity is the change in the electronic potential induced by atomic displacements associated with phonon modes described by wave vector $\mathbf{q}$, branch $\nu$, and frequency $\omega_{\mathbf{q}\nu}$. From these perturbations, one constructs the electron-phonon matrix elements $g_{mn\nu}(\mathbf{k},\mathbf{q})$, which describe the scattering amplitude between an initial electronic state $\epsilon_{n\mathbf{k}}$  and a final electronic state $\epsilon_{m\mathbf{k+q}}$ mediated by phonon $\omega_{\mathbf{q}\nu}$. Physically relevant quantities for superconductivity are then computed by enforcing energy conservation through delta functions centered at the initial and final electron states near the Fermi surface. When the Fermi energy is set to zero, 
the electron-phonon coupling strength is schematically given by: 

\[
\lambda_{\mathbf{q}\nu} \propto \int |g_{mn\nu}(\mathbf{k},\mathbf{q})|^2 \delta(\epsilon_{n\mathbf{k}})\delta(\epsilon_{m\mathbf{k+q}})d\mathbf{k},
\]
and the Eliashberg spectral function is:
\[
\alpha^2F(\omega) \propto \sum_{v} \int \omega_{\mathbf{q}\nu}\lambda_{\mathbf{q}\nu}\delta(\omega-\omega_{\mathbf{q}\nu})d\mathbf{q}. 
\]
The evaluation of these quantities involves summations over both electron $\mathbf{k}$ and phonon $\mathbf{q}$ states in their Brillouin zones, which in turn requires dense sampling to resolve the double-delta function accurately. In practice, this translates to evaluating e-ph matrix elements for a vast number of scattering processes involving electron pairs $(\mathbf{k}, \mathbf{k+q})$, making the calculation computationally expensive. To address this, several packages\,\cite{Zhou2021Perturbo,Lee2023,Li2024GWPT,Luo2024PRX} implement a Wannier interpolation scheme\,\cite{Giustino2007Wannier}: the matrix elements are first computed on a coarse grid using DFPT and then interpolated onto fine $\mathbf{k}$ and $\mathbf{q}$ grid meshes. While this approach enables high-resolution results, the construction of maximally localized Wannier functions can be challenging, especially for complex materials, and may not be easily automated for high-throughput studies.

In this work, we explore several strategies to accelerate the calculation of electron-phonon coupling interactions in the context of superconductivity calculations. First, we discuss a double-delta truncation method that significantly accelerates direct e-ph coupling calculations by one to two orders of magnitude. The key idea is to restrict both the initial and final electronic states to lie in the vicinity of the Fermi surface, thereby filtering out scattering pairs $(\mathbf{k},\mathbf{k+q})$ that do not contribute to the coupling process. To further improve efficiency and accuracy, we introduce non-uniform Farey grids\,\cite{chen2022}, constructed by combining uniform grids, which provide enhanced nonuniform resolution in specific regions of the Brillouin zone. These grids inform the exploitation of denser sampling where the e-ph scattering is strong, and coarser sampling where contributions are negligible, thereby optimizing computational cost. The same principle is applied to the phonon $\mathbf{q}$ grid to accelerate convergence of phonon Brillouin zone integrals. We demonstrate the effectiveness and robustness of our approach across a diverse set of materials, including elemental metals, binary compounds, and high-pressure hydrides, highlighting its general applicability and suitability for both accurate calculations of complex materials and efficient calculations in high-throughput studies.
%

\section{Double delta truncation}
\label{sec:double_delta}
The e-ph coupling strength $\lambda_{\mathbf{q}\nu}$ associated with a phonon mode $(\mathbf{q},\nu)$ can be written as:
\begin{equation}
    \lambda_{\mathbf{q}\nu} =  \frac{1}{N_F \omega_{\mathbf{q}\nu} }\sum_{mn} \int \frac{d\mathbf{k}}{\Omega_{\text{BZ}}}  |g_{mnv}(\mathbf{k},\mathbf{q})|^2 \delta(\epsilon_{n\mathbf{k}}) \delta(\epsilon_{m\mathbf{k}+\mathbf{q}}),
    \label{eq:lambda-definition}
\end{equation}
where $\omega_{\mathbf{q}\nu}$ is the phonon frequency, and $g_{mnv}(\mathbf{k},\mathbf{q})\propto \langle \psi_{m \mathbf{k}+\mathbf{q}} | \partial_{\mathbf{q}\nu} V | \psi_{n \mathbf{k}} \rangle$ is the e-ph matrix element describing the scattering amplitude from an initial state $|\psi_{n\mathbf{k}}\rangle$ of energy $\epsilon_{n\mathbf{k}}$ to a final state $|\psi_{m \mathbf{k}+\mathbf{q}}\rangle$ of energy $\epsilon_{m \mathbf{k}+\mathbf{q}}$, summed over all the bands and integrated over the Brillouin zone. The phonon energy in the second delta function $\delta(\epsilon_{m\mathbf{k}+\mathbf{q}})$ is neglected to work within the adiabatic approximation\,\cite{Giustino2017}. For simplicity, the Fermi level $\epsilon_F$ is set to zero.

The product of delta functions $\delta(\epsilon_{n\mathbf{k}}) \delta(\epsilon_{m\mathbf{k}+\mathbf{q}})$ enforces the energy conservation for both initial and final states. In practice, each $\delta$ function is approximated by a Gaussian function $\mathcal{G}_{\sigma}$ with a small smearing width $\sigma$, allowing for numerical integration: 
\begin{equation}
\delta(\epsilon_{n\mathbf{k}}) = \lim_{\sigma \to 0} \mathcal{G}_{\sigma}(\epsilon_{n\mathbf{k}})\approx \mathcal{G}_{\sigma}(\epsilon_{n\mathbf{k}}),
\end{equation}
where $\mathcal{G}_{\sigma}(\epsilon_{n\mathbf{k}}) = \frac{1}{\sqrt{2\pi}\sigma}\exp\!\left(-\frac{\epsilon_{n\mathbf{k}}^2}{2\sigma^2}\right)$.
The Gaussian is localised in a small region around the Fermi level, which motivates the definition of a top-hat function $\Theta_W(\epsilon_{n\mathbf{k}} )$ to truncate all electronic states falling outside a Fermi window of width $2W$ centred at the Fermi level:
\begin{equation}
\delta(\epsilon_{n\mathbf{k}}) \approx \mathcal{G}_{\sigma}(\epsilon_{n\mathbf{k}}) \approx \Theta_W(\epsilon_{n\mathbf{k}} )\,
\mathcal{G}_\sigma(\epsilon_{n\mathbf{k}} ),
\end{equation}
with:
\begin{equation}
\Theta_W(\epsilon_{n\mathbf{k}}) =
\begin{cases}
1, & |\epsilon_{n\mathbf{k}}| \le W, \\
0, & |\epsilon_{n\mathbf{k}}| > W.
\end{cases}
\end{equation}

This approximation can be used to reduce the number of electronic states that need to be explicitly included in the calculation. Specifically, the delta function product $\delta(\epsilon_{n\mathbf{k}}) \delta(\epsilon_{m\mathbf{k}+\mathbf{q}})$ can be approximated as $\mathcal{G}_{\sigma}(\epsilon_{n\mathbf{k}}) \mathcal{G}_{\sigma}(\epsilon_{m\mathbf{k}+\mathbf{q}})\Theta_W(\epsilon_{n\mathbf{k}})\Theta_W(\epsilon_{m\mathbf{k}+\mathbf{q}})$, which means that the only electronic states contributing to the integral in Eq.\,\eqref{eq:lambda-definition} are those for which $\Theta_W(\epsilon_{n\mathbf{k}})\Theta_W(\epsilon_{m\mathbf{k}+\mathbf{q}}) =1$. That is, only initial and final pairs of states that \textit{both} fall within the window $W$ need to be considered. In turn, this means that only those $\mathbf{k}$ points for which both the initial state and the final state have at least one band within the Fermi window contribute. 



We implement the double truncation strategy in \textsc{Quantum ESPRESSO}\,\cite{Giannozzi2009} as follows. First, the symmetry-reduced electronic eigenvalues $\epsilon_{n\mathbf{k}}$ are read from a non-self-consistent calculation performed on a dense $\mathbf{k}$-point grid. We then apply the single-delta truncation by evaluating $\Theta_W(\epsilon_{n\mathbf{k}})$ for all states and retaining only those $\mathbf{k}$ points that satisfy $\Theta_W(\epsilon_{n\mathbf{k}})=1$. The selected $\mathbf{k}$ points are subsequently unfolded to the full Brillouin zone using the crystal symmetry operations. For each irreducible $\mathbf{q}$ vector, a second truncation is applied by evaluating $\Theta_W(\epsilon_{m\mathbf{k}+\mathbf{q}})$ for the previously recovered $\mathbf{k}$ points that survived the single-delta truncation. This yields, for each irreducible $\mathbf{q}$, a set of $\mathbf{k}$ points satisfying the double-delta condition
\[
\Theta_W(\epsilon_{n\mathbf{k}})\,\Theta_W(\epsilon_{m\mathbf{k}+\mathbf{q}})=1.
\]
The resulting $\mathbf{k}$ list is then reduced to the irreducible first Brillouin zone using symmetry operations. This symmetry-reduced $\mathbf{k}$ list is finally passed to the DFPT code to perform the electron--phonon calculation for the corresponding irreducible $\mathbf{q}$ vectors only. In practice, the adoption of this double-delta truncation substantially decreases the number of electron-phonon coupling matrix elements $g(\mathbf{k},\mathbf{q})$ that must be explicitly evaluated.

\begin{figure}[ht]
    \centering
    \includegraphics[width=0.48\textwidth]{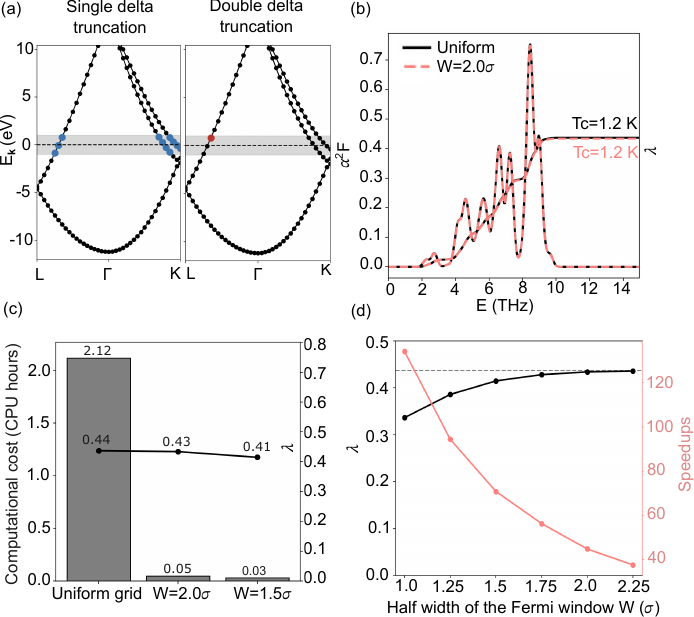}
    \caption{(a) Calculated band structure of aluminium with the Fermi level set to zero and the Fermi window indicated in grey. The left panel shows the $\mathbf{k}$ points selected by the single delta truncation $\Theta_W(\epsilon_{n\mathbf{k}}) =1$ (blue dots). The right panel illustrates the additional filtering imposed by the double delta truncation $\Theta_W(\epsilon_{n\mathbf{k}})\Theta_W(\epsilon_{m\mathbf{k}+\mathbf{q}}) =1$. For visual clarity, the shaded Fermi window is set to $W=1$\,eV in this illustration while in the calculations we use $W=100$\,meV.
    (b) Comparison of the Eliashberg spectral function $\alpha^2F$ and the e-ph coupling constant $\lambda$ computed using a full dense uniform $\mathbf{k}$-point grid ($84\times84\times84$, solid black lines) and the double delta truncated method (dashed red lines).
    (c) Computational cost comparison between the full uniform grid calculation and the double delta truncation method.
    (d) Convergence of $\lambda$ as a function of the half width of the Fermi window size $W$ used in the double delta truncation. The dashed horizontal line indicates the reference value from the uniform grid calculation.}
    \label{fig:Al} 
\end{figure}

Figure\,\ref{fig:Al}(a) shows the calculated band structure of aluminium, with the Fermi level set to zero and the Fermi window $W=100$\,meV in calculations. In the left panel, the single delta truncation $\Theta_W(\epsilon_{n\mathbf{k}}) =1$ selects all $\mathbf{k}$ points lying within the Fermi window, highlighted in blue. The right panel illustrates the double delta truncation $\Theta_W(\epsilon_{n\mathbf{k}})\Theta_W(\epsilon_{m\mathbf{k}+\mathbf{q}}) =1$, which filters the previously selected blue $\mathbf{k}$ points further, leaving only the red point relevant for the specific momentum transfer $\mathbf{q}=(1/4,0,0)$. Across the Brillouin zone, this reduces the number of required electron-phonon matrix elements from 204,680 to 2,538 for a grid of size $84\times84\times84$.

Figure\,\ref{fig:Al}(b) displays the converged Eliashberg spectral function $\alpha^{2}F$ and the integrated e-ph coupling constant $\lambda(\omega)$ (black solid lines) calculated using a dense uniform $\mathbf{k}$-point grid of $84\times84\times84$, which we take as the reference. Applying the double delta truncation to the uniform grid, only those $\mathbf{k}$ points satisfying $\Theta_W(\epsilon_{n\mathbf{k}})\Theta_W(\epsilon_{m\mathbf{k}+\mathbf{q}}) =1$ are included in the e-ph coupling evaluation. The results obtained using the truncated method are shown as dashed red lines in Figure\,\ref{fig:Al}(b), demonstrating excellent agreement with the full uniform grid calculation. All spectral peaks are accurately reproduced with the correct amplitudes using the double delta truncation. The e-ph coupling constant $\lambda$ also closely matches the uniform grid result, yielding a consistent predicted critical temperature of $T_{\mathrm{c}} = 1.2$\,K. This confirms that the double delta truncation effectively identifies the physically relevant $\mathbf{k}$ points for e-ph scattering, reducing the computational cost dramatically from 127 minutes to just 3 minutes, as shown in Figure\,\ref{fig:Al}(c).

A critical parameter in the double delta truncation is the half width of the Fermi window $W$, which defines the extent of the truncation. Figure\,\ref{fig:Al}(d) shows the variation of the computed e-ph coupling constant $\lambda$ as a function of $W$. The dashed horizontal line indicates the reference value obtained from the uniform grid. The $\lambda$ values obtained using truncation are systematically lower than the reference, highlighting the inherent underestimation introduced by truncation. However, $\lambda$ exhibits clear convergence with increasing Fermi window size, as more $\mathbf{k}$ points are included in the integration. This demonstrates the robustness of the double delta truncation, where accuracy is tunable via the Fermi window size. Remarkably, even $W=1.5\sigma$  yields a close approximation ($\lambda = 0.41$) to the reference value ($\lambda = 0.44$), with further improvement to $\lambda = 0.43$ at $W=2\sigma$, which can be considered converged. Although enlarging the Fermi window increases the runtime slightly from 2 to 3 minutes, the overall speedup remains substantial compared to the 127-minute cost of the full uniform grid calculation.

\begin{figure}[ht]
    \centering
    \includegraphics[width=0.48\textwidth]{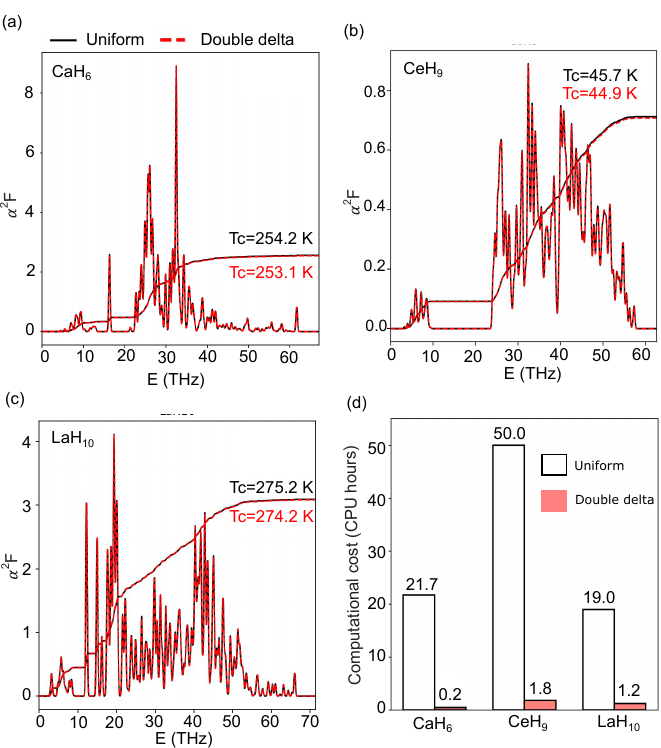}
    \caption{The calculated Eliashberg spectral function for hydrides (a) \ce{CaH6}, (b) \ce{CeH9} and (c) \ce{LaH10}. (d) The computational cost of using a uniform grid and a double delta truncation for each system.}
    \label{fig:hydride} 
\end{figure}

To demonstrate the applicability of the double delta truncation to more complex systems, Figure\,\ref{fig:hydride} presents the Eliashberg spectral function and e-ph coupling constant for high-pressure hydrides $\ce{CaH6}$, $\ce{CeH9}$, and $\ce{LaH10}$, computed using both a uniform grid and the double delta truncation. In Figure\,\ref{fig:hydride}(a), $\ce{CaH6}$ exhibits three prominent peaks in the Eliashberg spectrum at 16, 26, and $32$\,THz, respectively. The double delta truncation, shown as dashed red lines, successfully reproduces all three peaks, and the predicted critical temperature of $T_{\mathrm{c}} = 253.1$\,K agrees closely with the full grid result of $T_{\mathrm{c}} = 254.2$\,K. This suggests that the peak contributions originate entirely from $\mathbf{k}$ points within the double Fermi window, as the double delta function decays exponentially, demonstrating the universality of the truncation regardless of the detailed structure of the e-ph matrix elements $g(\mathbf{k},\mathbf{q})$.

Figure\,\ref{fig:hydride}(b) presents results for $\ce{CeH9}$, whose Eliashberg spectral function displays a continuous distribution from 25 to 50\,THz, indicating that many phonon modes contribute to the e-ph scattering. Even in this complex scattering regime, the double delta truncation performs remarkably well, accurately reproducing the spectral function, e-ph coupling constant, and $T_{\mathrm{c}}$. Figure\,\ref{fig:hydride}(c) shows results for $\ce{LaH10}$, which exhibits the highest $T_{\mathrm{c}}$ among these compounds. Compared to $\ce{CeH9}$, $\ce{LaH10}$ features pronounced individual peaks in the 10-20\,THz range and a continuous spectral distribution from 20 to 55\,THz. The double delta truncation successfully captures both the discrete peaks and the continuous distribution, yielding results in excellent agreement with the full grid calculation.

Figure\,\ref{fig:hydride}(d) compares the computational cost of the full grid calculation with that of the double delta truncation. For $\ce{CaH6}$, the computational cost is reduced from 21.7 to 0.2 CPU hours, and for $\ce{LaH10}$, from 19.0 to 1.2 CPU hours. In the case of $\ce{CeH9}$, whose primitive cell contains two formula units, the computational time is reduced from 50.0 CPU hours to just 1.8 CPU hours. Overall, the double delta truncation reduces the computational cost by one to two orders of magnitude, while maintaining high accuracy with an approximate 1\% error in $T_{\mathrm{c}}$ compared to the full grid calculation. These results establish the double delta truncation as an accurate and efficient approach for overcoming the high computational cost of e-ph coupling calculations in superconductors. 

Table\,\ref{tab:Tc_lambda_speedup} summarizes the performance of the double delta truncation method across a wide range of materials, including elemental metals, binary compounds, and high-pressure hydrides. The superconducting critical temperatures $T_{\mathrm{c}}$ computed using the truncated method show excellent agreement with those from full uniform grid calculations, with the largest absolute deviation being only 1.09\,K for \ce{CaH6}, corresponding to a relative error of just 0.43\%. Across all tested systems, the method achieves substantial computational gains, with speedup factors ranging from 16 to 105. This demonstrates the broad applicability and robustness of the double delta truncation scheme across diverse electronic and phononic environments. 


In addition, double-delta truncation is complementary to the density of states rescaling methodology developed in Ref.\,\cite{bozier_high-throughput_2026} by some of the present authors. DOS rescaling corrects the Eliashberg spectral function using an independently converged density of states at the Fermi level, enabling accurate $T_{\mathrm{c}}$ predictions at low cost when relatively coarse grids adequately describe the electron–phonon matrix elements and the dominant error arises from the DOS. Double-delta truncation makes dense-grid calculations more affordable by restricting the calculation to scattering channels within the Fermi window. These complementary strengths suggest a combined workflow in which DOS rescaling first enables low-cost screening across large sets of materials. For promising candidates, double-delta truncation could then enable finer sampling of the electron–phonon matrix elements to assess the convergence of the predicted $T_{\mathrm{c}}$. 

\begin{table*}[t]
  \centering
  \caption{
    Comparison of e-ph coupling constant $\lambda$ and transition temperature $T_{\mathrm{c}}$ (in K) computed using a full uniform grid and the double-delta truncation with a Fermi window half-width $W=2\sigma$. The computational cost using the double-delta truncation (in CPU time) and resulting speedup are also shown. 
  }
  \label{tab:Tc_lambda_speedup}

  \begin{ruledtabular}
    \begin{tabular}{
      l
      d{1.3}
      d{1.3}
      d{3.2}
      d{3.2}
      d{3.1}
      r
    }
      Material
      & \multicolumn{1}{c}{$\lambda$ (Uniform)}
      & \multicolumn{1}{c}{$\lambda$ (Truncated)}
      & \multicolumn{1}{c}{$T_{\mathrm{c}}$ (Uniform)}
      & \multicolumn{1}{c}{$T_{\mathrm{c}}$ (Truncated)}
      & \multicolumn{1}{c}{Cost (min)}
      & \multicolumn{1}{c}{Speedup} \\
      \midrule

      Al
      & 0.437 & 0.434
      & 1.22 & 1.17
      & 2.9 & 45 \\

      In
      & 0.949 & 0.943
      & 4.34 & 4.29
      & 13.7 & 26 \\

      $\beta$-Sn
      & 0.800 & 0.797
      & 3.75 & 3.72
      & 161.5 & 29 \\

      Pb
      & 2.058 & 2.045
      & 9.20 & 9.14
      & 9.5 & 28 \\

      TaC
      & 0.785 & 0.780
      & 10.30 & 10.15
      & 10.9 & 32 \\

      ZrN
      & 0.833 & 0.830
      & 15.77 & 15.65
      & 7.5 & 20 \\

      CaC$_6$
      & 0.864 & 0.864
      & 15.16 & 15.13
      & 67.5 & 16 \\

      CeH$_9$
      & 0.713 & 0.708
      & 45.70 & 44.90
      & 110.2 & 28 \\

      CaH$_6$
      & 2.565 & 2.551
      & 254.21 & 253.12
      & 12.2 & 105 \\

      LaH$_{10}$
      & 3.093 & 3.086
      & 275.22 & 274.23
      & 71.2 & 16 \\

    \end{tabular}
  \end{ruledtabular}
\end{table*}

\section{Non-uniform electronic grid}
To further enhance the efficiency of e-ph coupling calculations for high-throughput predictions of $T_{\mathrm{c}}$, we employ a non-uniform grid to better sample the Fermi surface. 
The intricate topology of the Fermi surface typically requires a dense uniform grid to achieve convergence. 
While the double delta truncation efficiently filters out $\mathbf{k}$ points outside the double Fermi window, those within the window remain uniformly distributed.
This suggests that introducing a non-uniform grid within the Fermi window can further improve sampling efficiency.

A key question is how to optimally construct such a non-uniform grid without prior knowledge of the system and without expensive pre-calculations. To achieve this, we approximate the electron-phonon matrix elements to be a constant  $g_{mn\nu}(\mathbf{k},\mathbf{q}) \approx g_{\nu}(\mathbf{q})$  and we can rewrite the expression for lambda $\lambda_{\mathbf{q}\nu}$
\[
\lambda_{\mathbf{q}\nu} =  \frac{|g_{\nu}(\mathbf{q})|^2}{N_F \omega_{\mathbf{q}\nu} }\sum_{mn} \int \frac{d\mathbf{k}}{\Omega_{\text{BZ}}}  \delta(\epsilon_{n\mathbf{k}}) \delta(\epsilon_{m\mathbf{k}+\mathbf{q}}).
\]
Within this approximation, the non-uniform grid is selected such that it reproduces the double delta integral obtained on a dense uniform mesh. This criterion enables grid construction without evaluating the full matrix elements $g_{mn}(\mathbf{k},\mathbf{q})$ , which typically constitute the dominant computational cost, while the double delta term can be computed at negligible expense.
Moreover, the double delta function decays exponentially and dominates the integrand.
Therefore, a selected non-uniform grid that accurately samples the double delta can also be used to perform e-ph coupling calculations with high fidelity.

\begin{figure}[ht]
    \centering
    \includegraphics[width=0.48\textwidth]{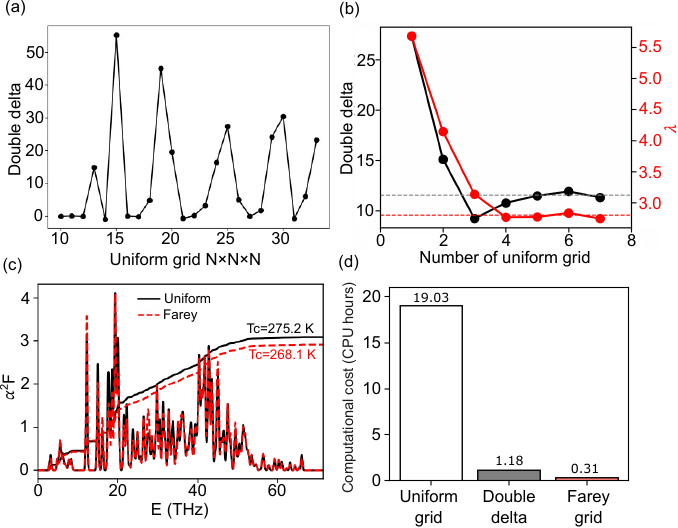}
    \caption{(a) Double delta values computed for \ce{LaH10} using various uniform $\mathbf{k}$-point grids.
    (b) Double delta values (black) and corresponding e-ph coupling constants $\lambda$ (red) computed using Farey grids formed by combining neighboring uniform grids. Reference results (dashed lines) are obtained using a uniform grid of size $48\times48\times48$.
    (c) Eliashberg spectral function $\alpha^2F$ and $\lambda$ for \ce{LaH10} computed using a dense uniform grid and a Farey grid. 
    (d) Computational cost comparison between a uniform grid, a double delta truncation on a uniform grid and a Farey grid.}
    \label{fig:farey-k} 
\end{figure}

Figure\,\ref{fig:farey-k}(a) illustrates the computed double delta values for \ce{LaH10} using various uniform grids (See SI for other cases).
The pronounced fluctuations underscore the difficulty of accurately sampling the Fermi surface.
For example, with a uniform grid of $15\times15\times15$, the double delta peaks at approximately 54\,eV$^{-2}$, but drops to zero for a grid of $16\times16\times16$.
With increasing grid density, another peak appears at $19\times19\times19$, followed by another drop to zero at $21\times21\times21$.
These fluctuations reflect the narrow width of the Gaussians used in the smearing function. 
Grids such as $15\times15\times15$ and $19\times19\times19$ happen to sample the narrow Gaussians, but this leads to overestimated results due to disproportionate $\mathbf{k}$-point weights. 
Conversely, grids such as $16\times16\times16$ and $21\times21\times21$ miss the narrow Gaussians entirely, leading to underestimated results.

A natural solution is to combine neighbouring uniform grids to form a Farey grid\,\cite{chen2022}, which improves sampling efficiency and convergence.
For instance, combining the $15\times15\times15$ and $16\times16\times16$ grids allows the former to sample the narrow Gaussians while the latter corrects the 
$\mathbf{k}$-point weights.
The advantage of the Farey grid lies in its efficiency: the coarse grid samples the Fermi surface, and the auxiliary grid does not significantly increase computational cost when truncation is applied.
In other words, the auxiliary grid contributes no 
$\mathbf{k}$ points within the double Fermi window but helps re-weight the primary grid to improve accuracy.

Figure\,\ref{fig:farey-k}(b) presents an example of the double delta computed using Farey grids formed by combining consecutive neighbouring uniform grids, using the uniform $25\times25\times25$ grid as a starting point.
For $n = 1$, the Farey grid reduces to a single uniform grid $25\times25\times25$, yielding a double delta of 27 eV$^{-2}$.
For $n = 2$, the addition of the $24\times24\times24$ grid reduces the double delta value to 15 eV$^{-2}$. 
As more grids are incorporated, the double delta values converge toward the reference result indicated by the dashed line.
To validate the Farey grid approach, full e-ph coupling calculations were also performed, shown as red curves in Figure\,\ref{fig:farey-k}(b).
The computed $\lambda$ is initially overestimated but gradually converges to the reference value (red dashed line), confirming that the Farey grid derived from the double-delta convergence also yields reliable sampling for the e-ph coupling constant.

Figure\,\ref{fig:farey-k}(c) compares the Eliashberg spectral function and e-ph coupling constant computed using a uniform grid and a non-uniform Farey grid. 
The Farey grid accurately captures the continuous spectrum from 20 to 50\,THz, though it slightly underestimates the amplitudes of individual peaks in the 10-20\,THz range.
The predicted $T_{\mathrm{c}}=268.1$\,K deviates by only 2.6\% from the reference value of 275.2\,K, indicating high accuracy. In practice, the choice of the initial starting grid size does not significantly influence the results as the Farey grid is able to find the optimal grids to match the reference double delta (See SI for results from various starting grids).
Remarkably, as shown in Fig.\,\ref{fig:farey-k}(d), the computational cost decreases from 1142 minutes to 71 minutes with the double delta truncation, and further to just 19 minutes with the Farey grid also applied, corresponding to a combined speedup by a factor of $60$.

\begin{figure}[ht]
    \centering
    \includegraphics[width=0.48\textwidth]{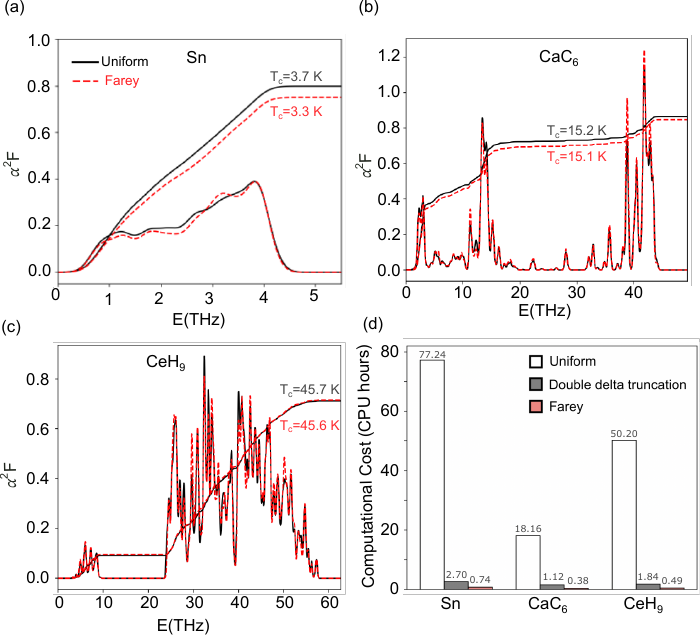}
    \caption{
    Comparison of the calculated Eliashberg spectral function $\alpha^2F$ and $\lambda$ for (a) \ce{Sn}, (b) \ce{CaC6} and (c) \ce{CeH9} using a dense uniform grid (black) and a Farey grid (red). 
    (d) The computational cost comparison in these three systems using a uniform grid, a double delta truncation on a uniform grid and a Farey grid.}
    \label{fig:farey-examples} 
\end{figure}

To further assess the applicability of the Farey grid, we consider three additional systems, $\beta$-\ce{Sn}, \ce{CaC6}, and \ce{CeH9}, which are the systems with largest computational cost listed in Table\,\ref{tab:Tc_lambda_speedup} after applying the double delta truncation (together with the already discussed LaH$_{10}$). As shown in Fig.\,\ref{fig:farey-examples}, the predicted superconducting transition temperatures $T_{\mathrm{c}}$ obtained using Farey grids exhibit excellent overall accuracy. Moreover, the Farey grid provides an additional speedup of approximately a factor of four compared to using only the double-delta truncation on a uniform dense grid. By comparing the results with our previous work on DOS rescaling, we find that both methods achieve similar accuracy at comparable computational cost (see Fig.\,S2 in the SI). Notably, even for a complex system such as \ce{CeH9}, the Farey grid substantially enhances sampling efficiency, reducing the computational cost from 50.2 CPU hours to 0.49 CPU hours and yielding an overall speedup of 102.

\section{Non-uniform phonon grid}
Previous examples have demonstrated how applying the Farey grid improves sampling over the electron $\mathbf{k}$ space. Convergence in the phonon $\mathbf{q}$ space can be equally challenging. Therefore, applying a Farey grid to $\mathbf{q}$ space can also enhance the accuracy of e-ph coupling calculations. 

\begin{figure}[ht]
    \centering
    \includegraphics[width=0.48\textwidth]{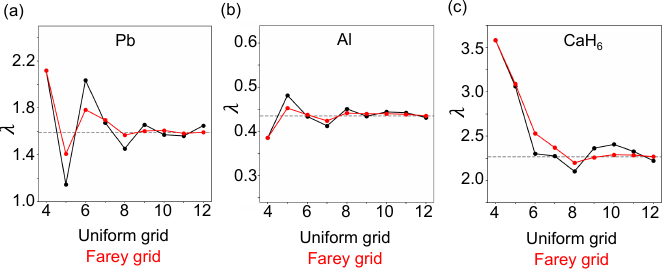}
    \caption{The calculated electron phonon coupling strength $\lambda$ using uniform grid (black) and Farey grid (red) for (a) Pb, (b) Al and (c) \ce{CaH6}. The x axis for a uniform grid represents its grid size while for a Farey grid, it represents a combination of previous uniform grids starting from the $4 \times 4 \times 4$ grid.}
    \label{fig:farey-q} 
\end{figure}

Figure\,\,\ref{fig:farey-q}(a) presents the calculated e-ph coupling constant $\lambda$ on a series of uniform $\mathbf{q}$ grids for Pb. Strong fluctuations are observed around the reference value $\lambda = 1.59$. For instance, a coarse uniform grid of $4 \times 4 \times 4$ yields a significantly overestimated $\lambda = 2.12$, largely due to an anomalously high value $\lambda_{\mathbf{q}} = 11.93$ at the point $\mathbf{q} = (0,-0.5,-0.5)$, which dominates the integral due to its large weight. Conversely, the $5 \times 5 \times 5$ grid underestimates $\lambda$ because it does not sample this critical $\mathbf{q}$-point, and similar inconsistencies continue to appear in subsequent grid sizes. These fluctuations arise from a few $\mathbf{q}$-points with exceptionally strong scattering strength being improperly weighted due to sparse and uniform sampling. A Farey grid, constructed by combining neighbouring uniform grids, offers a solution by redistributing sampling density to capture such features more effectively. 

Figure\,\,\ref{fig:farey-q}(a) also shows the $\lambda$ values computed using Farey grids, where the horizontal axis denotes the grid size for a uniform grid and the Farey combinations starting from the $4 \times 4 \times 4$ grid. For example, augmenting the $4 \times 4 \times 4$ grid with the $5 \times 5 \times 5$ grid introduces the point $\mathbf{q} = (0,-0.4,-0.4)$, which corrects the overestimated weight of $\mathbf{q} = (0,-0.5,-0.5)$, reducing it from 0.047 to 0.012 and consequently lowering $\lambda$ from 2.12 to 1.41. This more accurate representation of the Brillouin zone leads to smoother convergence with respect to the size of the phonon $\mathbf{q}$ grid.

Overall, Fig.\,\ref{fig:farey-q}(a) illustrates how for Pb, a uniform phonon $\mathbf{q}$ grid as dense as $12 \times 12 \times 12$ is not fully converged, while a Farey grid achieves convergence by combining uniform grids up to a size of $8 \times 8 \times 8$ only. The superior performance of the Farey grid in this system stems from its enhanced sampling resolution at the Brillouin zone boundaries, where critical scattering processes occur.

For systems with small variations in $\mathbf{q}$ space, the benefit of Farey grids is less dramatic but still appreciable. Figure\,\,\ref{fig:farey-q}(b) compares the convergence behaviour of $\lambda$ in aluminium using both uniform and Farey grids. The uniform grid initially overestimates $\lambda$ on a $4 \times 4 \times 4$ grid and underestimates $\lambda$ on a $5 \times 5 \times 5$ grid and it leads to converged results from a $6 \times 6 \times 6$ grid. The Farey grid $4 \times 4 \times 4 \cup 5 \times 5 \times 5$ yields better results than either a $4 \times 4 \times 4$  or a $5 \times 5 \times 5$ grid. Once a uniform grid reaches $6 \times 6 \times 6$, the Farey grid combining these grids also achieves convergence and performs only marginally better than a uniform grid. 

Finally, it is worth noting that Farey grids sample the region near the Brillouin-zone centre in the exact same manner as uniform grids\,\cite{chen2022}. For systems where e-ph coupling is dominated by phonon modes near the zone centre, Farey grids only offer limited benefits. Fig.\,\ref{fig:farey-q}(c) shows $\lambda$ of \ce{CaH6} as a function of the uniform and Farey grid sizes. The value of $\lambda$ is dramatically overestimated on a uniform $4 \times 4 \times 4$ grid due to an anomalously high value at $\mathbf{q} = \Gamma$ with $\lambda_{\mathbf{q}} = 51.5$ and its weight $\frac{1}{4^3}$. Increasing the grid density to $6\times 6 \times 6$ reduces the weight to $\frac{1}{6^3}$ and corrects $\lambda$ from 3.83 to 2.55, much closer to the reference value.  For example, the weight of the $\Gamma$ point on a Farey grid of $4\times4\times4\cup5\times5\times5\cup6\times6\times6$ is also $\frac{1}{6^3}$, the same as that of a uniform grid of size $6\times6\times6$. As a result, the Farey grid shows no improvement over the uniform grid until the grid $8\times 8 \times 8$, where it begins to reduce residual fluctuations in a manner similar to the aluminium case. It is important to highlight that the Farey grid does as well as the uniform grid in these zone-centre dominated systems, so it is still a generally valid strategy to use, for example, in high-throughput studies. However, this feature also indicates that alternative non-uniform strategies may be required to accelerate convergence for zone-centre-dominated systems. 

\section{Discussion}

The highly accurate results obtained using the double-delta truncation method demonstrate that only a small subset of electronic states near the joint Fermi surface significantly contribute to e-ph coupling. This finding suggests that the computational cost associated with traditional brute-force approaches, such as first-principles calculations on dense \(\mathbf{k}\)- and \(\mathbf{q}\)-grids or extensive Wannier interpolations, can be substantially reduced without compromising accuracy.
Our double-delta truncation scheme, combined with non-uniform electronic grids, provides a general and robust framework to isolate and evaluate only the physically relevant \((\mathbf{k}, \mathbf{k+q})\) scattering channels that meaningfully contribute to the coupling strength.
In addition, the non-uniform phonon grid captures the intricate geometry of the Fermi surface more effectively than uniform grids, enabling denser sampling in regions of strong scattering and coarser sampling elsewhere. This adaptivity not only improves sampling efficiency but also mitigates errors arising from highly localized coupling features in momentum space. More generally, this non-uniform sampling can be applied to other calculations requiring sampling the Brillouin zone, such as transport calculations and solving Bethe-Salpeter equations for excitonic properties.

\begin{figure}[ht]
    \centering
    \includegraphics[width=0.45\textwidth]{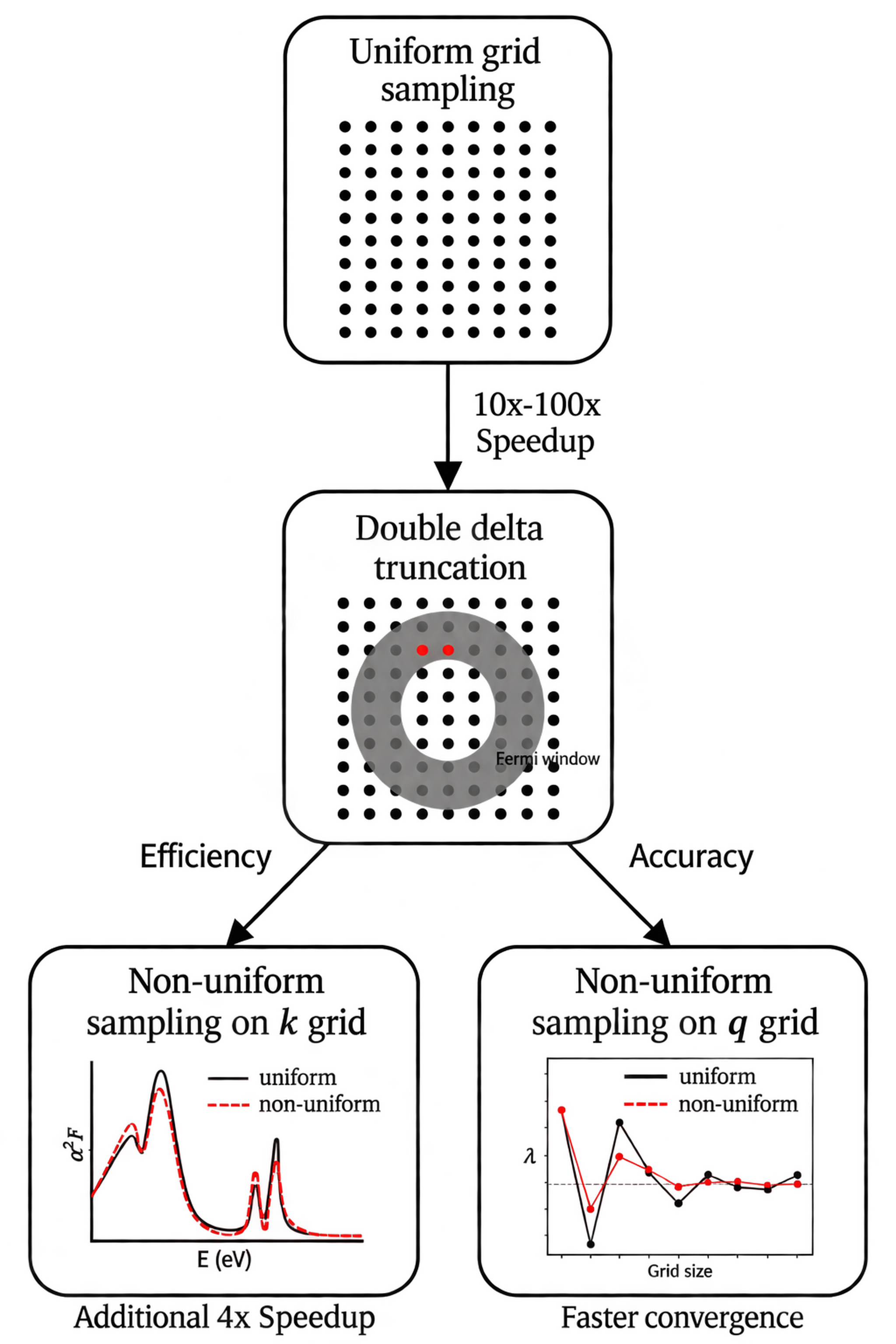}
    \caption{The workflow of applying the double delta truncation and the non-uniform sampling on $\mathbf{k}$ and $\mathbf{q}$ grids.}
    \label{fig:workflow} 
\end{figure}

Figure~\ref{fig:workflow} summarises the recommended workflow for combining these methods. Starting from a uniform electronic grid, double-delta truncation restricts the evaluation of electron–phonon matrix elements to scattering pairs whose initial and final states both lie within the chosen Fermi window, providing speedups of one to two orders of magnitude in the systems studied here. Further improvements can then be introduced through non-uniform sampling in electronic and phonon momentum space. For electronic sampling, a Farey grid selected using the convergence of the double-delta integral improves sampling efficiency and provides an additional speedup of approximately a factor of four in the examples considered. For phonon sampling, Farey grids can reduce fluctuations arising from strongly localised contributions in $\mathbf{q}$ space and improve convergence, with the benefit depending on the distribution of the dominant scattering processes. We therefore recommend applying double-delta truncation first, followed by non-uniform electronic sampling when further cost reduction is needed and non-uniform phonon sampling when convergence over momentum transfer remains challenging. These refinements can be applied separately or together. Finally, as discussed in Section~\ref{sec:double_delta}, this workflow can also complement DOS rescaling, enabling finer sampling for promising candidates identified through low-cost screening.

These methods have already been implemented within first-principles workflows based on \textsc{Quantum ESPRESSO}\,\cite{Giannozzi2009}, thereby demonstrating their practical integration in a widely used electronic structure package. Owing to their general formulation, they can also be incorporated into other frameworks such as \textsc{FHI-aims}\,\cite{blum_ab_2009}, \textsc{GPAW}\,\cite{mortensen_gpaw_2024}, \textsc{JDFTx}\,\cite{sundararaman_jdftx_2017} and \textsc{VASP}\,\cite{radevych_epw-vasp_2025}, making them broadly accessible to the computational materials science community.

\section{Conclusions}
In summary, we have shown that restricting both the initial and final electronic states to lie within a narrow window around the Fermi surface enables a dramatic reduction in computational cost. The proposed double-delta truncation method accelerates e-ph coupling calculations by one to two orders of magnitude, while maintaining high accuracy with errors typically below 1\% to 5\%. Furthermore, the use of non-uniform Farey grids offers a principled way to have a fine sampling in regions of the Brillouin zone that are most relevant to e-ph interactions. Taken together, these techniques pave the way for efficient and accurate large-scale or high-throughput predictions of phonon-mediated superconductivity and other complex superconducting materials.

\section{ACKNOWLEDGMENTS}
K.W., S.-W.K., and B.M. are supported by a UKRI Future Leaders Fellowship [MR/V023926/1]; S.C., C.J.P., and B.M. are supported by an EPSRC grant [EP/V062654/1]; and K.B. is supported by an EPSRC grant [EP/W524633/1]. S.-W.K. also acknowledges support from a Leverhulme Trust Early Career Fellowship [ECF-2024-052]. The computational resources were provided by the Cambridge Tier-2 system operated by the University of Cambridge Research Computing Service and funded by EPSRC [EP/P020259/1], and by the UK National Supercomputing Service ARCHER2, for which access was obtained via the UKCP consortium and funded by EPSRC [EP/X035891/1].
\section{Data availability}
The data that support the findings of this study are available within the paper and Supplementary Information. All other relevant data are available from the corresponding authors upon request.

\bibliography{ref}
\bibliographystyle{apsrev4-2}
\end{document}